\documentclass[prb,twocolumn,superscriptaddress,showpacs,amsmath,amssymb]{revtex4-1}

\usepackage{graphicx}
\usepackage{latexsym}
\usepackage{amsmath}
\usepackage{amssymb}
\usepackage{amsfonts}
\usepackage{changes}
\usepackage{color}
\usepackage{bm}% bold math
\usepackage{verbatim}
\usepackage{pagecolor,lipsum}
\usepackage{lineno}
\usepackage[version=3]{mhchem}

\usepackage{framed}
\definecolor{shadecolor}{RGB}{222,222,221}
\begin{document}

\title{Controllable magnetic states in chains of coupled $\varphi_0$-Josephson  junctions with ferromagnetic weak links}
%DOS and spin-resolved DOS in spin-textured S/F bilayers

\author{G.A. Bobkov}
\affiliation{Moscow Institute of Physics and Technology, Dolgoprudny, 141700 Russia}

\author{I. V. Bobkova}
\affiliation{Moscow Institute of Physics and Technology, Dolgoprudny, 141700 Russia}
\affiliation{National Research University Higher School of Economics, Moscow, 101000 Russia}

\author{A. M. Bobkov}
\affiliation{Moscow Institute of Physics and Technology, Dolgoprudny, 141700 Russia}

\date{\today}

 %%% abstract

\begin{abstract}
A  superconductor/ferromagnet/superconductor Josephson junction with anomalous phase shift ($\varphi_0$-S/F/S JJ) is a system, where the anomalous ground state shift $\varphi_0$ provides a direct magnetoelectric coupling between a magnetic moment and a phase of the superconducting condensate. If a chain of such $\varphi_0$-S/F/S JJs are coupled via superconducting leads, the condensate phase, being a macroscopic quantity, mediates a long-range interaction between the magnetic moments $\bm M_i$ of the weak links. We study static and dynamic magnetic properties of such a system. It is shown that it manifests properties of $n$-level system, where the energies of the levels are only determined by projections of the total magnetic moment $\sum \bm M_i$ onto the easy magnetic axis. It is similar to  a magnetic atom in a Zeeman field, but the role of the field is played by the magnetoelectric coupling. However, unlike an atom in a magnetic field, the relative order of energies of different states is controlled by electrical means. It is also demonstrated that  $\sum \bm M_i$ can be fully controlled by a supercurrent and the response of the magnetic system to local external perturbations is highly nonlocal.
\end{abstract}

 %%% PACS numbers
  \maketitle
 
\section{Introduction}

Physics of equilibrium magnetic states and magnetic excitations crucially depends on the type of magnetic interaction between the magnetic moments. Direct and indirect exchange interactions, spin-orbit effects and dipole-dipole interactions lead to a large number of different magnetic states, including ferromagnetism, antiferromagnetism, altermagnetism, helimagnetism, skyrmions, spin glasses, etc. Another interesting direction is an interaction of more macroscopic magnets, which is very important, for example, for magnetoresistive phenomena\cite{Baibich1988,Binasch1989}. It can be realized via indirect interlayer exchange interaction \cite{Grunberg1986,Majkrzak1986,Salamon1986}, dipole-dipole interaction and, what is more close to the subject of the present paper, also via superconductors. 

The coupling via superconductors can be realized by different physical mechanisms. One of them is the proximity effect. As it was first pointed out by de Gennes, a superconductor makes the antiferromagnetic configuration of magnets more favorable \cite{deGennes1966}. The reason is that with such a mutual orientation of magnets, superconductivity in the interlayer is less suppressed as a result of partial compensation
of paramagnetic depairing. The characteristic scale of such an interaction is the superconducting coherence length $\xi_S$. There are a lot of theoretical proposals and experimental realizations of a superconducting spin valve basing on the interaction via the proximity effect \cite{Tagirov1999,Aristov1997,Leksin2011,Li2013,DiBernardo2019,Ghanbari2021,Zhu2017,Koshelev2019}. Later it was proposed \cite{Devizorova2019} that an interaction between magnets can also be mediated by  the electromagnetic proximity effect \cite{Mironov2018}, the essence of which is the appearance of Meissner currents in a superconductor in response to the presence of an adjacent magnetic material. The characteristic scale of this coupling is the penetration depth of the magnetic field $\lambda$.  

Recently another mechanism for establishing interaction between magnetic moments has been proposed. It is not related to the proximity effects  and is based on another physical principle. The interaction is of magnetoelectric origin and is mediated by supercurrents, what makes it extremely long-range and  decaying according to a power law. The magnetoelectric coupling was considered both for magnetic impurities in superconductors \cite{Malshukov2018,Lu2023,Xiang2023} and for  weak links of coupled Josephson junctions (JJs) \cite{Bobkov2022}. In the last case the interaction is mediated by the phase of a superconducting condensate, which is a macroscopic quantity. For this reason the characteristic scale of the interaction is not restricted by the typical proximity scales of a superconductor, such as $\xi_S$ and $\lambda$, and can be much larger \cite{Bobkov2022}. 

The coupling between the superconducting phase and the magnetic moment is realized in JJs with a strong spin-orbit coupling in the interlayer region \cite{Krive2004,Nesterov2016,Reynoso2008,Buzdin2008,Zazunov2009,Brunetti2013,Yokoyama2014,Bergeret2015,Campagnano2015,Konschelle2015,Kuzmanovski2016,Malshukov2010} or in JJs on a
topological insulator \cite{Tanaka2009,Linder2010,Zyuzin2016,Lu2015,Dolcini2015}, where the surface conduction electrons have the property of the full spin-momentum locking \cite{Burkov2010,Culcer2010,Yazyev2010,Li2014}.
Physically the presence of the coupling between the superconducting phase and the magnetic moment in JJs manifests itself in the form of the so-called anomalous phase shift in the ground state of the junction \cite{Bobkova_review,Shukrinov_review}. The essence of this effect is that under the simultaneous breaking of inversion symmetry, which allows for the spin-orbit coupling (SOC), and a time-reversal symmetry, which is due to the presence of the magnetic moment, a supercurrent can be induced in the JJ at zero phase difference between the leads.  In the ground state of
the junction this ”anomalous supercurrent” is compensated by the phase shift $\varphi_0 \neq 0,\pi$, which is called the anomalous ground state phase shift and the JJs
manifesting this effect are called $\varphi_0$-JJs.

The breaking of the time-reversal symmetry can be achieved more easily by applying a magnetic field to the JJ. JJs with anomalous phase shift generated by the Zeeman effect of the applied magnetic field have already been implemented experimentally by several groups \cite{Mayer2020,Szombati2016,Assouline2019,Murani2017}, including those on a topological insulator. Realization of $\varphi_0$-superconductor/ferromagnet/superconductor JJs ($\varphi_0$-S/F/S JJs) is a more challenging problem, but the possibility to obtain in such systems a direct coupling between the magnetization of the weak link and the superconducting phase opens great prospects for applications of such structures for controlling magnetization \cite{Konschelle2009,Shukrinov2017,Nashaat2019,Rabinovich2019,Guarcello2020,Bobkova2020,Bobkova_review}. One of the possibilities is to use for the interlayers two-dimensional (2D) or quasi 2D ferromagnets, where the Rashba spin-orbit coupling can be strong due to the structural inversion symmetry breaking. The other way is to exploit ferromagnetic insulator/three-dimensional topological insulator (3D TI) hybrids as interlayers \cite{Chang2013,Kou2013,Kou2013_2,Chang2015,Jiang2014,Wei2013,Jiang2015,Jiang2016}.

In Ref.~\onlinecite{Bobkov2022} it was shown that in a system of two coupled $\varphi_0$-JJs with ferromagnetic weak links characterized by magnetic moments $\bm M_{1,2}$ the magnetic state of the system, that is the directions of the both magnetizations can be fully controlled by the phase difference between the external superconducting leads. It was found that at large values of the phase difference the most favorable state of the magnets is ferromagnetic and the directions of the both magnetizations are dictated by the phase. At the same time at small values of the phase difference the most favorable state is antiferromagnetic. In this case the state of a given magnet is not fixed by the phase and, therefore, there is an indirect interaction between them, which is mediated by the superconducting phase. 

In the present paper we continue investigations of coupled $\varphi_0$-S/F/S JJs and generalize results of Ref.~\onlinecite{Bobkov2022} to the case of arbitrary number of coupled JJs. It is found that the system of arbitrary number of easy-axis magnets, which are weak links of the $\varphi_0$-JJs, in some aspects behaves like a magnetic atom. States with different projections of the total magnetic moment $\sum \bm M_i$ onto the easy axis (where sum is taken over all weak links) are degenerate in the absence of interaction mediated by the condensate phase, i.e. above the critical temperature of the superconductor or in the absence of the anomalous phase shift. Including the interaction removes this degeneracy, similar to the case of a magnetic atom in a Zeeman field. However, unlike an atom in a magnetic field, the relative energies of different projections of the total  magnetic moment are controlled by the external phase difference.

The paper is organized as follows. In Sec.~\ref{model} we describe the system  and the model, which we study. Sec.~\ref{equilibrium} is devoted to investigation of the equilibrium magnetic state of the system. In Sec.~\ref{dynamics} we study the dynamical processes of transition of the system between different stable states. In Sec.~\ref{non-equivalence} we discuss the influence of possible non-equivalence of different JJs on the effects considered above. Our conclusions are presented in Sec.~\ref{conclusions}.

\section{Model}
\label{model}

\begin{figure}[tb]
	\begin{center}
		\includegraphics[width=75mm]{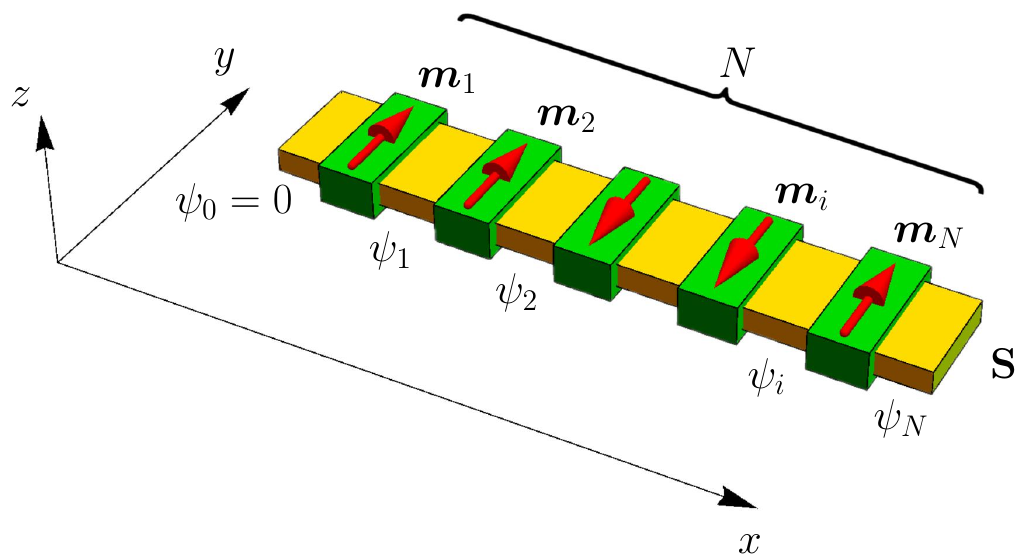}
		\caption{Sketch of a coupled system of $N$ $\varphi_0$-S/F/S JJs.  Magnetic moment of each JJ is shown by a red arrow. $\psi_i$ is a phase of the superconducting region connecting $i$-th and $i+1$-th weak links with respect to the phase of the left lead, which is taken to be zero.}
  \label{fig:sketch}
	\end{center}
 \end{figure}

We consider a linear chain of $N$ coupled $\varphi_0$-S/F/S JJs, where S means a conventional superconductor and F means a homogeneous ferromagnet with magnetic moment $\bm M_i$, where $i$ is a number of the weak link in the chain. Further we introduce unit vectors along the direction of the corresponding magnetization $\bm m_i = \bm M_i/|\bm M_i|$. It is assumed that the ferromagnets are easy-axis magnets with the easy axis along the $y$-direction. The sketch of the system is represented in Fig.~\ref{fig:sketch}. The superconducting phase difference $\psi_N$ between the leads is an external controlling parameter. 

The physical origin of the inversion symmetry breaking leading to the $\varphi_0$-behavior can be different and does not influence our conclusions. For example, one can use for the interlayers a few monolayer van der Waals ferromagnets up to the monolayer limit \cite{Gibertini2019,Ai2021,Kang2022}, where the Rashba SOC can be intrinsic or due to the structural inversion symmetry breaking. The other possibility is a complex interlayer made of a combination of a thin-film ferromagnet and a heavy metal layer like Pt, providing strong SOC. The interlayer can be also composed of  a ferromagnetic insulator on top of the 3D  TI, as it is mentioned in the introduction.  If the ferromagnet is an insulator, it is assumed that the magnetization $\bm M$ of the ferromagnet induces an effective exchange field $\bm h \sim \bm M $ in the underlying conductive layer.

The current-phase relation (CPR) of a separate S/F/S JJ takes the form $I = I_{c,i} \sin (\chi_i-\varphi_{0,i})$, where $I_{c,i}$ is the critical current of $i$-th magnet, $\chi_i = \psi_i - \psi_{i-1}$ is the superconducting phase difference at this JJ and $\varphi_{0,i}$ is the anomalous phase shift for a given JJ. In general, anomalous phase shift $\varphi_{0,i}$ in S/F/S JJs depends on the direction of magnetization $\bm m_i$. The particular form of this dependence is determined by the type of spin-orbit coupling. For definiteness we consider Rashba type SOC because it arises due to the structural inversion asymmetry and is the most common type of SOC for low-dimentional ferromagnets and thin-film ferromagnet/normal metal (F/N) hybrid structures. In this case the anomalous phase shift takes the form 
\begin{eqnarray}
\varphi_0 =  r \hat {\bm  j} \cdot (\bm n \times \bm m ),
\label{phi_0}
\end{eqnarray}
where $\hat {\bm  j}$ is the unit vector along the Josephson current and $\bm n$ is the unit vector describing the direction of the structural asymmetry in the system. For the case under consideration it is along the $z$-axis. $r$ is a constant quantifying the strength of the coupling between the magnetic moment and the condensate. It depends on the material parameters of the ferromagnet, Rashba constant $\alpha$, length of the ferromagnetic interlayer and was calculated in the framework of different models \cite{Buzdin2008,Bergeret2015}. Eq.~(\ref{phi_0}) is also valid for the S/F/S JJs on top of the 3D TI, where it has been predicted that $r = 2hd/v_F$ \cite{Zyuzin2016,Nashaat2019}, where $d$ is the length of the interlayer and $v_F$ is the Fermi velocity of the surface conduction electrons in 3D TI. The results presented below depend only on the symmetry of Eq.~(\ref{phi_0}) expressing how the anomalous phase shift depends on the direction of the magnetization $\bm m_i$, and the dependence of the constant $r$ on the junction parameters is irrelevant for our conclusions.
If we choose $x$-axis along the Josephson current, then symmetry of our system dictates that
\begin{eqnarray}
\varphi_{0,i} = r_i m_{yi}.
\label{chi0}
\end{eqnarray}
This relation also survives in the dynamic situation $\bm m_i = \bm m_i(t)$ and has been used  for calculation of the magnetization dynamics in voltage-biased and current-biased JJs \cite{Nashaat2019,Konschelle2009,Shukrinov2017,Guarcello2020}. Now it is clear that our choice of the magnetic easy axis along the $y$-direction maximizes the magnetoelectric coupling between the magnetic moment and the superconducting phase.

In the framework of our model we assume that the critical current $I_{c,i}$ does not depend on the direction of the magnetization $\bm m_i$. In fact, the behavior of the critical current depends crucially on the particular type of the considered S/F/S JJ. For example, it can be independent on the magnetization direction, as it has been reported for the ferromagnets with SOC \cite{Buzdin2008}, or it can depend strongly on  the $x$-component of the magnetization, as it takes place for the ferromagnetic interlayers on top of the 3D TI \cite{Zyuzin2016,Nashaat2019}. For the case of two coupled JJs the influence of the dependence $I_c(\bm m)$ on the results was considered in Ref.~\onlinecite{Bobkov2022}. It was found that taking into account this dependence does not change the results qualitatively, but just modify boundaries of different regimes.

The energy of the system consists of Josephson energies of  all junctions and easy-axis anisotropy energies of  all magnets:
\begin{eqnarray}
E =  \sum \limits_{i=1}^N \Bigl[ \frac{\hbar I_{c,i}}{2e}\bigl(1-\cos(\psi_i-\psi_{i-1}-\varphi_{0,i})\bigr) - \nonumber \\ 
\frac{K_i V_{F,i}}{2} m_{yi}^2 \Bigr],
\label{energy}
\end{eqnarray}
where the first term is the Josephson energy and the second term is the magnetic anisotropy energy. $K_i$ - is the anisotropy constant of $i$-th magnet and $V_{F,i}$ is its volume.
$\psi_{i}$ is a phase of $i$-th superconductor (see  Fig.~\ref{fig:sketch}). The current conservation dictates 
\begin{eqnarray}
I_{c,i}\sin(\psi_i-\psi_{i-1}-\varphi_{0,i}) = 
I_{c,j}\sin(\psi_j-\psi_{j-1}-\varphi_{0,j}) 
\label{current_conservation}
\end{eqnarray}
for arbitrary $i$ and $j$. 

In general, one should take into account the phase gradient due to the supercurrent flowing through the system. This leads to the fact that the phase $\psi_i$ of $i$-th superconductor is not constant, $\psi_i(x) = \psi_{i,l} + \kappa_i (I_{chain}/I_{c,i})(x/L)$, where $\psi_{i,l}$ is the superconducting phase at the left end of $i$-th superconductor, $L$ is its length and the second term accounts for the phase gradient due to the supercurrent $I_{chain}$ flowing through the system. The coefficient $\kappa_i$ quantifies the relation between the superconducting phase gradient and the supercurrent and can be estimated as $\kappa_i \sim e I_{c,i} L/\sigma_S \Delta S$, where $\Delta$ is the superconducting order parameter, $\sigma_S$ is the normal state conductivity of the $i$-th superconductor and $S$ is its cross section. To simplify the analysis we disregard the order parameter phase gradient. Its influence on the phase diagram of the coupled system was investigated in Ref.~\onlinecite{Bobkov2022} for the case of two JJs and it was found that it results in a renormalization of the parameter $r$ and does not influence qualitatively the results. In addition, numerical values of $\kappa_i$ were estimated for realistic systems and it was concluded that the phase gradient can be safely neglected at least up to submillimeter lengths of the superconductors \cite{Bobkov2022}.

 \begin{figure}[tb]
	\begin{center}
		\includegraphics[width=65mm]{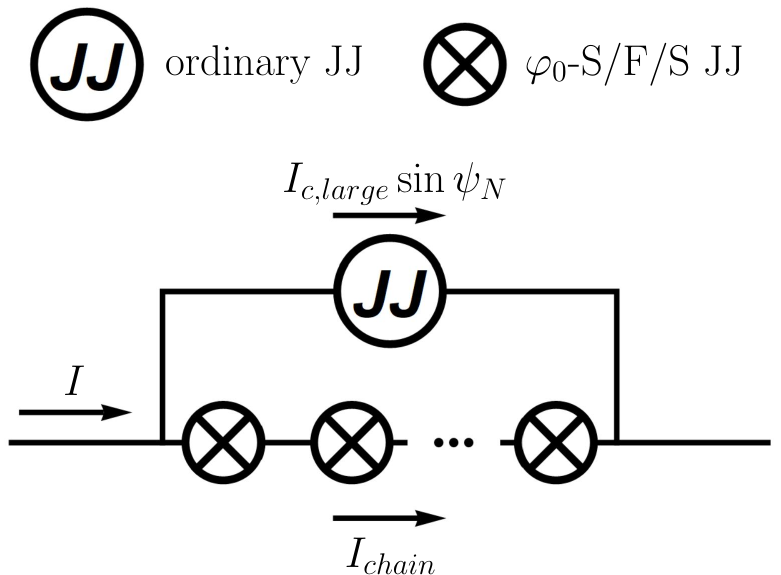}
		\caption{Sketch of an asymmetric Josephson interferometer, where the coupled system of $\varphi_0$-JJs is in parallel with an ordinary JJ with large critical current $I_{c,large} \gg I_{c,i}$. The scheme allows to control phase $\psi_N$ via the external current $I$ (see text).}
  \label{fig:sketch_2}
	\end{center}
 \end{figure}

For the problem under consideration it is important to have a fixed phase difference $\psi_N$ between the external superconducting leads and to have a possibility to control it. Experimentally the phase $\psi_N$ can be controlled by several ways. One of them is to insert the considered system into the superconducting loop under the applied magnetic flux.  However, the
direct proportionality between the Josephson phase and magnetic flux in a
loop configuration only holds if the loop inductance is negligible.  
If it is non-negligible, peculiar effects may occur in the presence of $\varphi_0$-JJs \cite{Guarcello2020_SQUID}, which require additional care. The other way is to insert it into the asymmetric Josephson interferometer, where the considered system is in parallel with an ordinary JJ with a much higher critical current. Then the magnetic state of the system can be controlled by the external current. 
 In the present paper we consider the second way to control external phase difference. The scheme of the corresponding system is presented in Fig.~\ref{fig:sketch_2}. The total external current $I$ and the  phase $\psi_N$ between the external superconducting leads are related as
\begin{eqnarray}
I = I_{c,large}\sin \psi_N + I_{chain}, 
\label{current_psi}
\end{eqnarray}
where $I_{c,large}$ is the critical current of the additional ordinary JJ and $I_{chain}$ is the current through the chain of the $\varphi_0$-S/F/S JJs. We assume that $I_{c,large} \gg I_{c,i}$. In this case $\psi_N \approx \arcsin[I/I_{c,large}]$. All the calculations of the dynamics of our coupled $\varphi_0$-S/F/S JJs, discussed in Sec.~\ref{dynamics}, are performed for the system sketched in Fig.~\ref{fig:sketch_2}, where we set the external current $I$.

\section{Phase-controlled equilibrium magnetic state}
\label{equilibrium}

At first we assume that all the coupled JJs are identical, that is they all have the same parameters $r$, $I_c$ and $K V_{F}$. The influence of variations of these parameters is studied below in Sec.~\ref{non-equivalence}. Then from Eq.~(\ref{current_conservation}) it follows that there are two possible solutions for the phase distribution along the chain of the coupled JJs. The first solution is that the total phase difference $\psi_i - \psi_{i-1} - \varphi_{0,i}$ at each of the JJs equals to $\Phi+2 \pi n_i$, where $n_i$ is an integer number. $\Phi$ can be found from the condition
\begin{eqnarray}
\sum \limits_{i=1}^N (\Phi + 2\pi n_i) = \sum \limits_{i=1}^N (\psi_i - \psi_{i-1} - \varphi_{0,i}),
\label{Phi_eq}
\end{eqnarray}
which gives us that
\begin{eqnarray}
\Phi = \frac{\psi_N}{N} - \frac{\sum \limits_{i=1}^N \varphi_{0,i}}{N} + \frac{2\pi n}{N},
\label{Phi_sol}
\end{eqnarray}
where $n = \sum \limits_{i=1}^N n_i$ - an integer number. Substituting this solution for the phase distribution into the Josephson energy, we obtain the following expression for the total energy of the system :
\begin{eqnarray}
E = N E_J (1-\cos \Phi) - 
E_M \sum \limits_{i=1}^N m_{yi}^2, 
\label{energy_1}
\end{eqnarray}
where $E_J = \hbar I_c/2e$ and $E_M = K V_F/2$.

The second solution of Eq.~(\ref{current_conservation}) is given by the total phase difference $\psi_i-\psi_{i-1}-\varphi_{0,i} = \pi - \Phi_u - 2 \pi n_i$ at $M$ JJs and $\psi_i-\psi_{i-1}-\varphi_{0,i} = \Phi_u + 2 \pi n_i$ at $N-M$ remaining JJs. In this case analogously to derivation of Eq.~(\ref{Phi_sol}) we obtain:
\begin{eqnarray}
\Phi_u = \frac{\psi_N - \sum \limits_{i=1}^N \varphi_{0,i} + 2\pi n - \pi M}{N-2M}, 
\label{Phi_sol2}
\end{eqnarray}
where $M<N/2$. The total energy of the system takes the form:
\begin{eqnarray}
E = E_J (N-(N-2M)\cos \Phi_u) - 
E_M \sum \limits_{i=1}^N m_{yi}^2. 
\label{energy_2}
\end{eqnarray}

For a given magnetic configuration, that is for a given values $m_{yi}$ at all weak links, the total energy of the system $E(\psi_N)$, described by Eq.~(\ref{energy_1}), as a function of the phase difference between the external superconducting leads $\psi_N$ has $N$ different branches. The total energy described by Eq.~(\ref{energy_2}) has $N-2M$ branches for a given $M$. Our numerical analysis shows that all the energy branches described by Eq.~(\ref{energy_2}) are unstable, that is for a given $\psi_N$ the system cannot live at such a branch and immediately goes to one of the branches described by Eq.~(\ref{energy_1}).

Further, using Eq.~(\ref{energy_1}) one can investigate which magnetic configurations of the system can be stable in the system for a given set of parameters $E_J$, $E_M$ and $r$. Due to the fact that the magnetic energy of the system is a concave function of $m_{yi}$ and the Josephson energy only depends on the magnetizations in the combination $\sum \limits_{i=1}^N \varphi_{0,i} = r\sum \limits_{i=1}^N m_{yi}$ the minima of $E$ as a function of $\left\{ m_{yi} \right\}$ can be only located at the edges of the hypercube build on $\left\{ m_{yi} \right\}$. That is, the minima  can only be at points, where no more that  one $|m_{yi}| \neq 1$. Then we can find a minimum of the energy with respect to all $m_{yi}$ independently.

Under the condition ${\rm sgn}[m_{yi}]\partial E/ \partial m_{yi}|_{m_{yi}=\pm 1}<0$ "corner states" $m_{yi} = \pm 1$ are  minima of the energy. This condition is fulfilled irrespective of the external phase difference $\psi_N$ if $2E_M/E_J >r$.  This regime is not interesting if our goal is to control the magnetic configuration via the external phase. For this reason in our study we focus on the opposite regime $2E_M/E_J <r$, when there is a solution
\begin{eqnarray}
\sin \Phi = 2E_M/rE_J ,
\label{instability_cond}
\end{eqnarray}
corresponding to the condition $\partial E/ \partial m_{yi} = 0$, at which the magnetic configuration, corresponding to a given energy branch, becomes unstable. This parameter region allows us to control the magnetic configuration by adjusting the external phase.

There is also another important condition, which determines the behavior of the magnetic configuration. One can define parameter regions, where only magnetic configurations corresponding to "corner states" $m_{yi} = \pm 1$ are stable and the parameters regions, where $m_{yi} \neq \pm 1$ ("non-aligned states") are possible. To find the corresponding parameter regions we have to analyze when $|m_{yi}| \neq 1$ can be a minimum of the energy. For this purpose we  consider conditions $\partial E/ \partial m_{yi} = -2E_M m_{yi} - E_J r \sin \Phi = 0$ and $\partial^2 E/ \partial m_{yi}^2 = E_J r^2 \cos \Phi/N - 2E_M >0$. From these conditions we find that the most favorable conditions for the realization of a state $m_{yi} \neq \pm 1$ are at $\cos \Phi = 1$, that is $\sin \Phi=0$ and $m_{yi} = 0$. This situation can be realized if 
\begin{eqnarray}
\frac{2E_M}{E_J}<\frac{r^2}{N}.
\label{nonaligned_cond}
\end{eqnarray}

\begin{figure}[tb]
	\begin{center}
		\includegraphics[width=60mm]{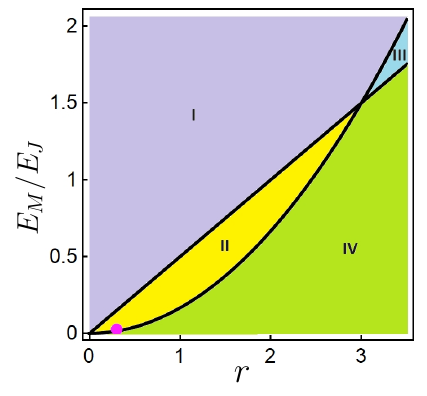}
		\caption{Phase diagram of the coupled $\varphi_0$-S/F/S JJs. The lines $2E_M/E_J = r$ and $2E_M/E_J = r^2/N$ divide the diagram into Regions I, II, III, IV. For description of the regions see text. Pink point indicates  parameters $(r,E_M/E_J)$ for which static and dynamic results presented in Figs.~\ref{fig:branches}-\ref{fig:reversal_3} are calculated.}
  \label{fig:phase_diagram}
	\end{center}
 \end{figure}

The above considerations can be summarized in the form of a phase diagram, which is represented in Fig.~\ref{fig:phase_diagram}. The lines $2E_M/E_J = r$ and $2E_M/E_J = r^2/N$ divide the diagram into four regions. In Region I "non-aligned states" are not allowed and all the "corner states" are always stable for an arbitrary $\Psi_N$. In Region II "non-aligned states" are not allowed, but "corner states" can become unstable for some phase difference. In Region III "non-aligned states" are allowed and all "corner states" are stable for an arbitrary $\Psi_N$. In Region IV  "non-aligned states" are allowed and all magnetic configurations become unstable for some value of $\Psi_N$. This phase diagram is a generalization of the phase diagram discussed in  Ref.~\onlinecite{Bobkov2022} to the case of arbitrary number of $\varphi_0$-S/F/S JJs $N \geq 2$.

\begin{figure}[tb]
	\begin{center}
		\includegraphics[width=85mm]{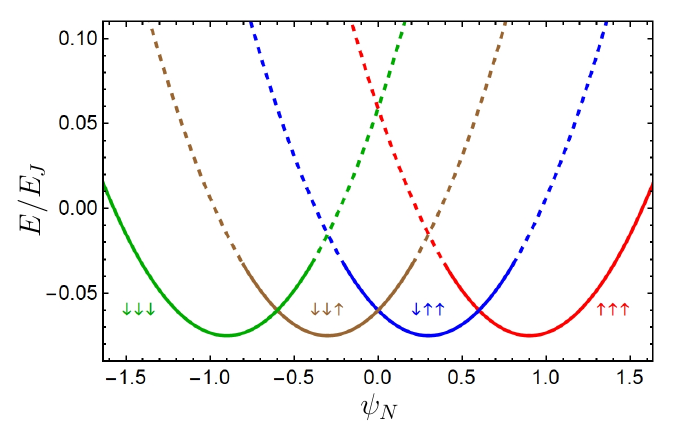}
		\caption{Lowest energy branches of $E(\psi_N)$ for $N=3$. There are 4 possible magnetic configurations. They are shown by arrows and for each of them the lowest energy branch is plotted by the corresponding color. Dashed parts of the curves represent parts of the corresponding branch, where the magnetic configuration becomes unstable. $r=0.3$, $E_M/E_J = 0.025$.}
  \label{fig:branches}
	\end{center}
 \end{figure}

Parameters $E_M/E_J$ and $r$ were estimated in Ref.~\onlinecite{Bobkov2022} for the model of the insulating ferromagnet on top of the 3D TI. If one takes the parameters corresponding to $Nb/Bi_2Te_3/Nb$ Josephson junctions \cite{Veldhorst2012}: the junction length $d=50 nm$, $I_{c} = 40 A/m$, $v_F = 10^5 m/s$ and assumes $K \sim [(10-10^2)erg/cm^3]\times d_F$  for yttrium iron garnet (YIG) thin films \cite{Mendil2019}, where $d_F = 10 nm$ is the F thickness along the $z$-direction, then one obtain $E_M/E_J \sim 10^{-2}-10^{-1}$. From the other hand, if for the ferromagnetic weak link one considers permalloy with very weak anisotropy, one can estimate $K \sim 10^3 erg/cm^3 \times d_F$ \cite{Konschelle2009}. It gives us $E_M/E_J >1$.  Therefore, different regimes from $E_M/E_J \ll 1$ to $E_M/E_J > 1$ can be realized experimentally. Basing on the experimental data on the Curie temperature of the magnetized TI surface states \cite{Jiang2015}, where the Curie temperature in the range $20-150K$ was reported, we can roughly estimate $h \lesssim 0.01- 0.1 h_{YIG}$ for YIG/3D TI interlayers. It corresponds to the dimensionless parameter $r = 2 h d/v_F \lesssim 2-13$. From the other hand, $r$ should be much smaller for ferromagnetic interlayers with intrinsic SOC or for combined interlayers consisting of a ferromagnet and a heavy metal because of the reducing factor $\Delta_{so}/\varepsilon_F$ \cite{Buzdin2008}, which is typically considerably less than unity. Here $\Delta_{so}$ is a typical value of the SOC-induced splitting of the electron spectra and $\varepsilon_F$ is the Fermi energy of the material. From these estimates it follows that, in principle, all regions of the phase diagram can be experimentally accessible.

Further we choose parameters of the system belonging to Region II (pink point in Fig.~\ref{fig:phase_diagram}) and investigate equilibrium and dynamic behavior of the system for the chosen set of parameters. This set of parameters is chosen because of (i) the possibility to control magnetic configuration by the external phase, (ii) smallness of typical realistic values of $E_M/E_J$ \cite{Bobkov2022} (except for ferromagnets with very weak magnetic anisotropy) and (iii) absence of the "non-aligned states", which complicate the physical picture. According to Eqs.~(\ref{chi0}),(\ref{Phi_sol}) and (\ref{energy_1}) the Josephson energy only depends on the total projection $M_y = \sum \limits_{i=1}^N m_{yi}$ of all the magnets on the easy axis. For this reason in the absence of the "non-aligned states" there are only $N+1$ essentially different magnetic configurations corresponding to $M_y = \left\{ -N, -N+2, ...,N-2, N \right\}$. For each of the configurations $E(\psi_N)$ has $N$ branches, which differ by value of $\Phi$, as described above and quantified by Eqs.~(\ref{Phi_sol}) and (\ref{energy_1}).

In Fig.~\ref{fig:branches} we demonstrate low-energy branches of $E(\psi_N)$ for $N=3$. There are 4 possible magnetic configurations and for each of the configurations the lowest energy branch is plotted. The other branches are not shown because they do not participate in the processes of dynamical switching between the magnetic configurations, described below. Dashed parts of the curves represent parts of the corresponding branch, where the magnetic configuration becomes unstable. The distance from the minimum of the corresponding branch, determined by the condition $\sin \Phi = 0$, to the point of instability, where the branch becomes dashed, according to Eq.~(\ref{instability_cond}) is determined as
\begin{eqnarray}
\Delta \psi_N = N \arcsin \left[ \frac{2 E_M}{rE_J} \right]
\label{instability_2}
\end{eqnarray}

Therefore, the analysis of the energy of the chain of coupled $\varphi_0$-S/F/S JJs indicates that the system behaves similarly to an atom, where all magnetic configurations are degenerate at $r=0$, but nonzero $r$ removes the degeneracy. The split states are characterized by different projections of the total magnetization $M_y$ of the macroscopic atom on the easy axis. But, unlike an atom in a Zeeman field, the order in energy of these split states can be changed by varying the external superconducting phase $\psi_N$, as it is illustrated in Fig.~~\ref{fig:branches}.

\section{Dynamics and switching of the magnetic state}
\label{dynamics}

Here we discuss several dynamical effects in the system of the coupled $\varphi_0$-S/F/S JJs, which directly illustrate the analogy between the system and the macroscopic $n$-level system. The dynamics of $i$-th magnet is described by the Landau-Lifshitz-Gilbert (LLG) equation \cite{Landau1935ONTT,Gilbert2004,Lakshmanan2011}: 
\begin{eqnarray}
\frac{\partial\bm m_i}{\partial t} = -\gamma \bm m_i \times \bm H_{eff} + \alpha \bm m_i \times \frac{\partial\bm m_i}{\partial t} - \nonumber \\
\frac{\gamma r I_{chain}}{2e M d d_F}[\bm m \times \bm e_y],~~~~~~
\label{LLG}
\end{eqnarray}
where $\gamma$ is the gyromagnetic ratio,  $\bm H_{eff} = (K/M) m_y \bm e_y$ is the local effective field in the ferromagnet induced by the easy-axis magnetic anisotropy and $\alpha$ is the Gilbert damping constant. The last term in Eq.~(\ref{LLG}) describes the spin-orbit torque, exerted on the magnet by the electric current $I_{chain}$  \cite{Yokoyama2011,Miron2010,Bobkova2018,Bobkova2020,Bobkov2022}. The torque is averaged over the ferromagnet thickness $d_F$ along the $z$-direction. 
The total current flowing through each of the JJs consists of the supercurrent and the normal quasiparticle current contributions \cite{Rabinovich2019}:
\begin{eqnarray}
I_{chain}=I_c \sin (\psi_i - \psi_{i-1} - \varphi_{0,i})+ \nonumber \\
\frac{1}{2eR_N}(\dot \psi_i - \dot \psi_{i-1} - \dot \varphi_{0,i}),
\label{current_total}
\end{eqnarray}
where $R_N$ is the normal state resistance of a separate $\varphi_0$-S/F/S JJ. The dynamics of all magnetizations $\bm m_{i}$ is calculated numerically from Eqs.~(\ref{LLG}), (\ref{current_total}) and (\ref{current_psi}). 

\begin{figure}[tb]
	\begin{center}
		\includegraphics[width=65mm]{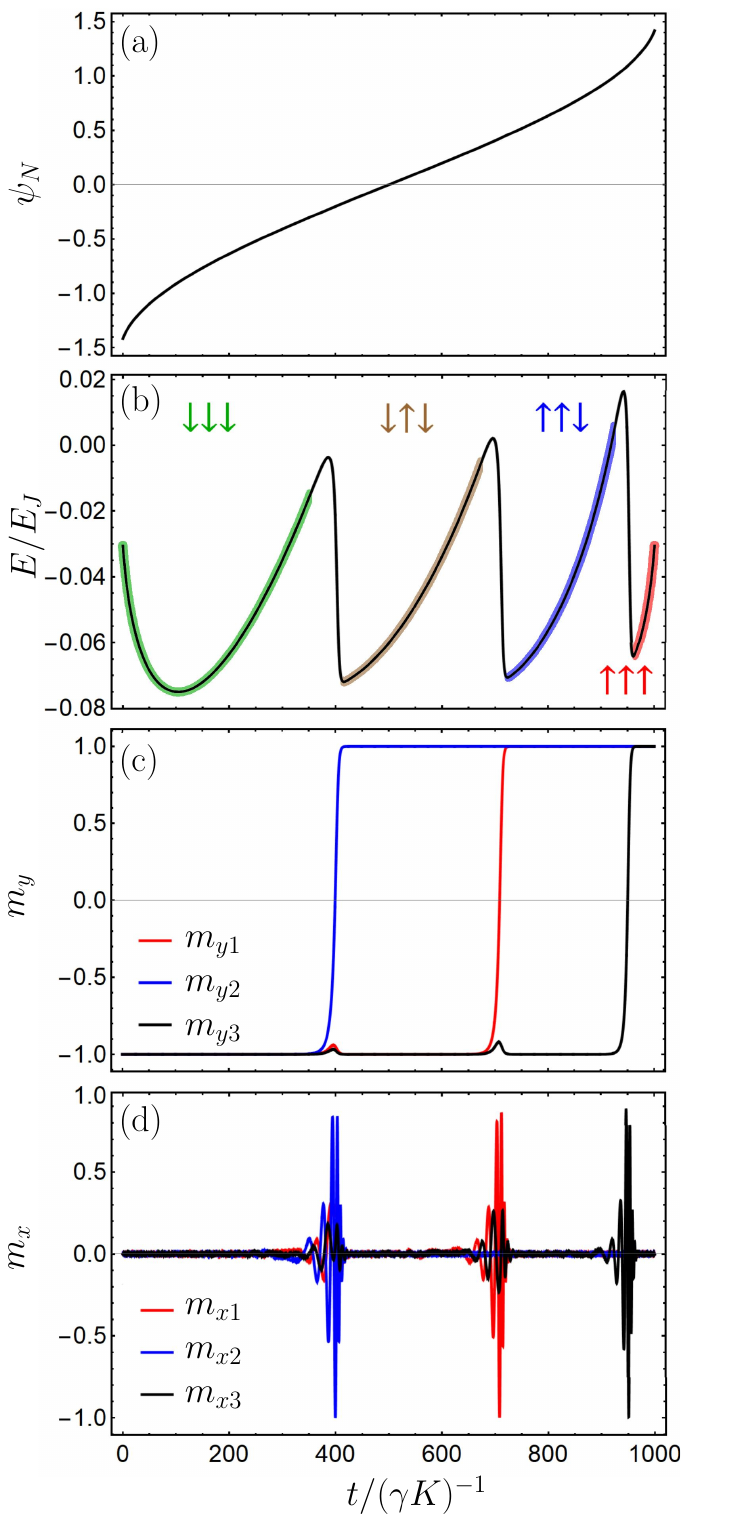}
		\caption{Time evolution of the energy and magnetic configuration upon varying the external current from $-I_{c,large}$ to $I_{c,large}$ according to $I=(\beta t-1) I_{c,large}$ with $\beta=0.002 (\gamma K)$. (a) External phase $\psi_N (t)$. (b) Energy of the coupled chain of $\varphi_0$-S/F/S JJs as a function of time. Intervals of $t$, where the system is in one of its stationary states, are marked with the colors coding the corresponding state in Fig.~\ref{fig:branches}. (c) $m_{yi}(t)$ and (d) $m_{xi}(t)$ for all magnets. $r=0.3,~\frac{E_M}{E_J}=0.025$, $\alpha=0.1$, $I_{c,large}/I_c=100$}
  \label{fig:dynamics_1}
	\end{center}
 \end{figure}

First of all, we demonstrate that in the setup sketched in Fig.~\ref{fig:sketch_2} any magnetic configuration can be realized by varying the external current $I $. Thus, the magnetic configuration is fully controllable by electric means. The results are presented in Fig.~\ref{fig:dynamics_1}. Fig.~\ref{fig:dynamics_1}(a) is auxiliary and demonstrates the dependence of the external phase difference on time when we vary the external current $I \propto t$. Fig.~\ref{fig:dynamics_1}(b) represents the dependence of the total energy of the system on time. In some ranges of $t$ the system is in its stationary states, shown in Fig.~\ref{fig:branches}. These time intervals are additionally marked with the corresponding colors, which coincide with colors used in Fig.~\ref{fig:branches}. Figs.~\ref{fig:dynamics_1}(c)-(d) illustrate the time evolution of all magnetic moments. $m_{yi}(t)$ are presented in Fig.~\ref{fig:dynamics_1}(c) and $m_{xi}(t)$ are plotted in Fig.~\ref{fig:dynamics_1}(d). $m_{zi}(t)$ are not shown because they behave very similar to $m_{xi}(t)$. 

Further we study the dynamics of the system caused by an external reversal of one of the magnets. To reverse one of the magnets we fix a definite value of the external current $I$, which corresponds to a particular stable magnetic configuration and apply a magnetic field $\bm H= \pm 4K \bm e_y$ during the time interval $\Delta t = 50 (\gamma K)^{-1}$ to one of the magnets. The resulting dynamics can be very different. Since the physics of the magnetic system is not determined by a magnetic interaction with any characteristic spatial scale, the reversal of a given magnet can cause reversals of any magnets in the system. The dynamics is only determined by the order and stability of energy levels of the system for a given phase.

\begin{figure}[tb]
	\begin{center}
		\includegraphics[width=65mm]{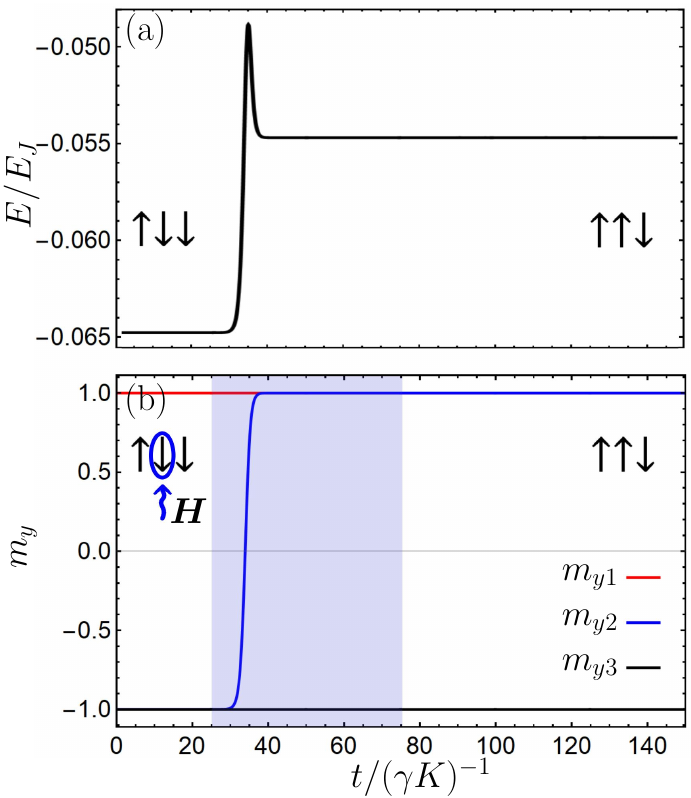}
		\caption{Time evotuion of (a) the system energy and (b) easy-axis projections of all magnets  $m_{yi}$ under external reversal of $\bm m_2$ by a pulse of an applied magnetic field.  Magnetic field $\bm H=4K \bm e_y$ is applied during the time interval shown by the colored region in panel (b).  $r=0.3,~\frac{E_M}{E_J}=0.025,~\alpha=0.1,~\psi_N=-0.05$.}
  \label{fig:reversal_2}
	\end{center}
 \end{figure}

In Fig.~\ref{fig:reversal_2} we demonstrate that if the initial magnetic configuration is stable and for a given phase difference there is another stable state corresponding to a higher energy, it is possible to excite the system to this state. We reverse one of the magnets, the resulting state is stable and the system remains in this state. In Fig.~\ref{fig:reversal_2}(a) we plot the dependence of the total energy of the system on time, and Fig.~\ref{fig:reversal_2}(b) represents the time evolution of all  magnetic moments. 

\begin{figure}[tb]
	\begin{center}
		\includegraphics[width=65mm]{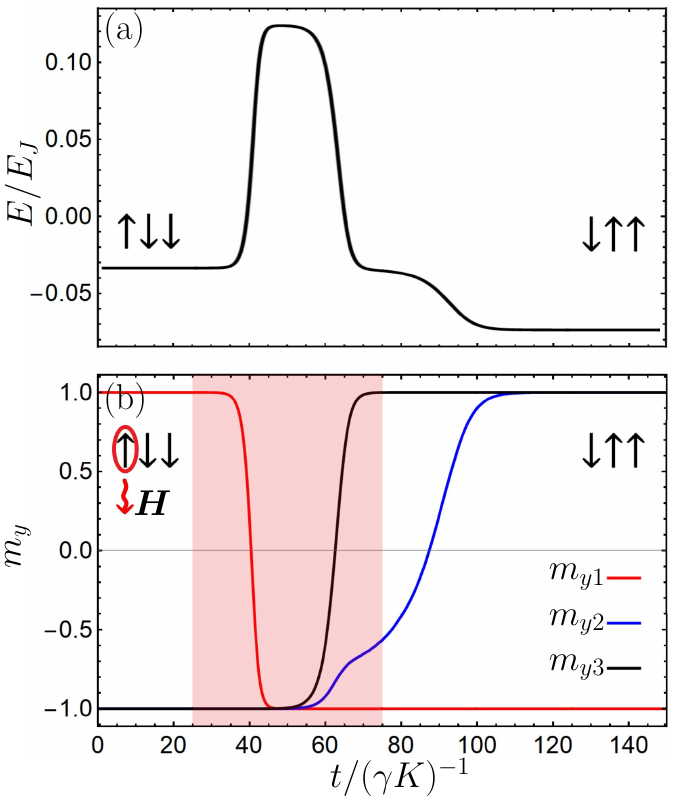}
		\caption{The same as in Fig.~\ref{fig:reversal_2} but for the different external phase difference $\psi_N$. Magnetic field $\bm H=-4K \bm e_y$ is applied to $\bm m_1$ during the time interval shown by the colored region in panel (b). $r=0.3,~\frac{E_M}{E_J}=0.025,~\alpha=0.1,~\psi_N=0.20$}
  \label{fig:reversal_1}
	\end{center}
 \end{figure}

\begin{figure}[tb]
	\begin{center}
		\includegraphics[width=65mm]{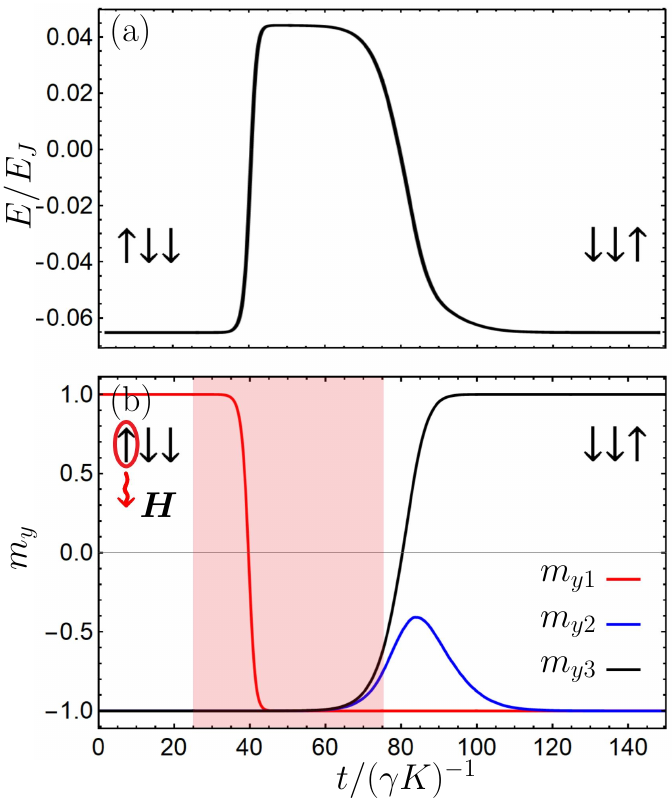}
		\caption{The same as in Fig.~\ref{fig:reversal_2} but for the different external phase difference $\psi_N$. Magnetic field $\bm H=-4K \bm e_y$ is applied to $\bm m_1$ during the time interval shown by the colored region in panel (b). $r=0.3,~\frac{E_M}{E_J}=0.025,~\alpha=0.1,~\psi_N=-0.05$}
  \label{fig:reversal_3}
	\end{center}
 \end{figure}

In Figs.~\ref{fig:reversal_1} and ~\ref{fig:reversal_3} we demonstrate the case when the initial magnetic configuration is stable for the chosen phase difference, but if we reverse one of the magnets, there is no stable state for the resulting magnetic configuration. Then different possibilities can be realized. The first option is shown in Fig.~\ref{fig:reversal_1}. The reversal of $\bm m_1$ by the magnetic field pulse leads to the reversal of all the other magnets. As a result the system switches to the lower energy state. 

The other possibility is illustrated in Fig.~\ref{fig:reversal_3}. We reverse $\bm m_1$ by the external magnetic field. For the chosen phase difference there is no stable state for the resulting configuration $\downarrow \downarrow \downarrow$ and, therefore, $\bm m_3$ is also reversed. But the probability to have $\bm m_2$ reversed is the same.  In this case, a random magnet may be reversed, it is not determined by the distance from the externally reversed magnet.

Figs.~\ref{fig:reversal_2}-\ref{fig:reversal_3} demonstrate the results for $N=3$ just as an example to illustrate the general behavior. If the chain consists of $N>3$ $\varphi_0$-S/F/S JJs, then at a given $\psi_N$ one can have more than two stable branches. Analogous increase of the number of stable branches (up to $N+1$) can also happen if one change the value of the parameter $E_M/E_J$ without changing $N$. In this case by reversing one of the magnets we can reach any of these stable configurations at a given $\psi_N$. That is, the physics is the same as in Figs.~\ref{fig:reversal_2}-\ref{fig:reversal_3}, but there are more possible initial and final states of the system.

\section{Effects of non-equivalence of JJs}
\label{non-equivalence}

Now we discuss how fluctuations of parameters $K_i$, $I_{c,i}$ and $r_i$, describing individual $\varphi_0$-S/F/S JJs, affect the results obtained above. We restrict ourselves by the parameter regions I and II of the phase diagram, where the "non-aligned states" do not occur. 
First of all, in this case variations of the magnetic anisotropy constant $K_i$ do not influence $E(\psi_N)$. This is due to the fact that at $m_{yi} = \pm 1$ the magnetic anisotropy energy only depends on the sum $\sum \limits_{i=1}^N K_i$.

Now let us consider the influence of variations of the critical currents of individual JJs $I_{c,i}$. Small variations of the critical current also does not influence $E(\psi_N)$. Indeed, we can write $I_{c,i} = I_{c,0}(1+x_i)$, where $\sum \limits_{i=1}^N x_i= 0$, that is $I_{c,0}$ is the average critical current of all the JJs. We assume $x_i \ll 1$. Let us define an auxiliary parameter $\Phi_0$ as $I_{chain} = I_{c,0}\sin \Phi_0$. Then $\Phi_i \equiv \psi_i - \psi_{i-1} - \varphi_{0,i} $ can be written as $\Phi_i \approx \Phi_0 + \delta \Phi_i$. From the condition $I_{chain} = I_{c,i} \sin \Phi_i = I_{c,0} \sin \Phi_0 $ up to the first order with respect to $x_i$ we obtain $\delta \Phi_i \approx -x_i \tan \Phi_0$. Then $\sum \limits_{i=1}^N \delta \Phi_i = 0$. From the conditions $\sum \limits_{i=1}^N \delta \Phi_i = 0$ and $\sum \limits_{i=1}^N  x_i = 0$ it immediately follows that the first-order term in the expansion of $E(\psi_N)$ on $x_i$ vanishes. Accounting for higher-order terms with respect to $x_i$ can result in asymmetry and distortions of the energy branches.

\begin{figure}[tb]
	\begin{center}
		\includegraphics[width=85mm]{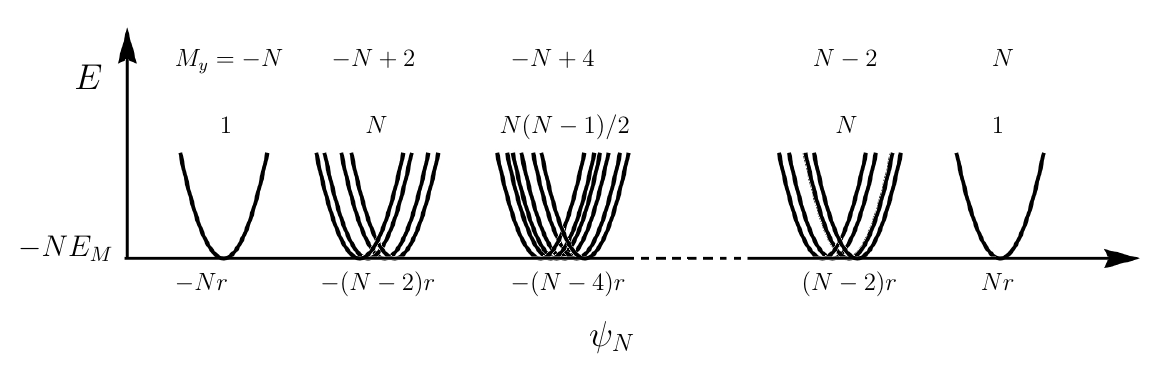}
		\caption{Schematic illustration of the splitting of the lowest energy branches, corresponding to different $M_y$. It is assumed that $\delta r_{ij} = |r_i - r_j| \ll \sum \limits_{i=1}^N r_i/N$, otherwise the splitting of each of the branches can become large and curves originated from different initial configurations can be mixed. It does not modify the result qualitatively, but makes it difficult to visually perceive the picture.}
  \label{fig:splitting}
	\end{center}
 \end{figure}

The most essential effect on the physics of the $N$-level system, discussed in this work, can be caused by variations of $r$. They result in the additional splitting of the energy branches. If $r$ is the same for all $\varphi_0$-S/F/S JJs, states $\{m_{y1},m_{y2}, ....\} = \{+1,-1, ....\}$ and $\{-1,+1, ....\}$ are degenerate. However, if $r_1 \neq r_2$, according to Eq.~(\ref{Phi_sol}) they have different values of $\Phi$ and, therefore different energies. 
In general, if all $r_i$ are different, an energy branch corresponding to $|M_y| = |N-2K|$ is split over $C_N^K$ branches.  It is illustrated in Fig.~\ref{fig:splitting}, where the lowest energy branchs for each of the magnetic configurations are shown.

In addition at high enough temperatures close to the critical temperature and for JJs with small critical currents $I_{c,i}$ the system could be also sensitive to thermal
fluctuations. The general stability of a single $\varphi_0$-JJ against the thermal fluctuations has been already studied \cite{Guarcello2020,Guarcello2021,Guarcello2023}. In our case of a chain of $\varphi_0$-JJs at a given $\psi_N$ there are several stable states in the system. For this reason the thermal fluctuations  will primarily spoil the stability of higher energy metastable states. To minimize the destructive effect of thermal fluctuations, it is necessary to work at low temperatures and  select larger JJs with large critical currents.

\section{Conclusions}
\label{conclusions}

In conclusion, we have studied static and dynamic magnetic properties of a system of $N$ coupled $\varphi_0$-S/F/S Josephson junctions, where the anomalous ground state phase shift $\varphi_0$ provides a direct coupling between magnetic moments and the phase of the superconducting condensate. The condensate phase, being a macroscopic quantity, mediates a long-range interaction between the magnetic moments. Due to this magnetoelectric coupling the system exhibits properties of $n$-level system, where the corresponding energies are only determined by different projections of the total magnetic moment $\sum \bm M_i$ onto the easy axis, similar to  a magnetic atom in a Zeeman field. However, unlike an atom in a magnetic field, the relative energies of different projections of the system's magnetic moment are controlled by the external phase difference. Further we demonstrate that if one inserts the coupled chain of JJs into an asymmetric SQUID, one can reach any of the states, corresponding to different projections of the total magnetic moment $\sum \bm M_i$ onto the easy axis, by varying the external current. It is also demonstrated that the long-range coupling between the weak links leads to highly nonlocal and nontrivial dynamics of the magnetic configuration under application of a local external perturbation to one of the weak links. The dynamics
is only determined by the order and stability of energy
levels of the system for a given phase and, therefore, further supports analogy with macroscopic $n$-level system. 

\begin{acknowledgments}
The analysis of equilibrium properties of the system has been supported by MIPT via Project
FSMG-2023-0014. The studies of the dynamics have been supported by RSF project No. 22-42-04408. 
\end{acknowledgments}

\bibliography{many_magnets}

\end{document}